\documentclass[apjl]{emulateapj}
\usepackage{times}
\usepackage{apjfonts}
\usepackage{amssymb}
\usepackage{epsfig}

\usepackage{graphicx,graphics}
\usepackage{dcolumn}
\usepackage{natbib}
\usepackage{times}
\def\kmm#1  {{\bf [KMM:~ #1]~}}
\def\new#1 {{\bf #1 }}
\def\cut#1 {\sout{#1} }

\newcommand{\hi}{H{\sc i}}
\newcommand{\ci}{C{\sc i}}

\newcommand{\dal}{\ensuremath{\lsb \Delta \alpha/ \alpha \rsb}}

\newcommand{\dmu}{\ensuremath{\lsb \Delta \mu/\mu \rsb}}
\newcommand{\beq}{\begin{equation}}
\newcommand{\eeq}{\end{equation}}

\newcommand{\lsb}{\left[}
\newcommand{\rsb}{\right]}

\shorttitle{Probing fundamental constant evolution with neutral atomic gas lines}
\shortauthors{Kanekar et al.}

\begin{document}

\title{Probing fundamental constant evolution with neutral atomic gas lines}

\author{N. Kanekar\altaffilmark{1,2},
J. X. Prochaska\altaffilmark{3},
S. L. Ellison\altaffilmark{4},
J. N. Chengalur\altaffilmark{1}
}

\altaffiltext{1}{National Centre for Radio Astrophysics, TIFR, Ganeshkhind, Pune - 411007, India}
\altaffiltext{2}{National Radio Astronomy Observatory, 1003 Lopezville Road, Socorro, NM87801, USA; nkanekar@nrao.edu}
\altaffiltext{3}{UCO/Lick Observatory, UC Santa Cruz, Santa Cruz, CA 95064, USA}
\altaffiltext{4}{Department of Physics and Astronomy, University of Victoria, Victoria, B.C., V8P 1A1, Canada}

\begin{abstract}

We have detected narrow \hi~21cm and \ci\ absorption at $z \sim 1.4 - 1.6$ towards 
Q0458$-$020 and Q2337$-$011, and use these lines to test for possible changes 
in the fine structure constant $\alpha$, the proton-electron mass ratio $\mu$, 
and the proton gyromagnetic ratio $g_p$. A comparison between the 
\hi~21cm and \ci\ line redshifts yields $\Delta X/X = [+6.8 \pm 1.0] \times 10^{-6}$ 
over $0 < \langle z \rangle < 1.46$, where $X = g_p \alpha^2/\mu$, and the errors 
are purely statistical, from the gaussian fits. The simple line profiles and the 
high sensitivity of the spectra imply that statistical errors in this comparison 
are an order of magnitude lower than in previous studies. Further, the \ci\ lines arise in 
cold neutral gas that also gives rise to \hi~21cm absorption, and both background quasars are 
core-dominated, reducing the likelihood of systematic errors due to local velocity offsets 
between the hyperfine and resonance lines. The dominant source of systematic error 
lies in the absolute wavelength calibration of the optical spectra, which appears uncertain to 
$\sim 2$~km/s, yielding a maximum error in $\Delta X/X$ of $\sim 6.7 \times 10^{-6}$. 
Including this, we obtain $\Delta X/X = [+6.8 \pm 1.0 (statistical) \pm 6.7 (max. systematic)] 
\times 10^{-6}$ over $0 < \langle z \rangle < 1.46$. Using literature constraints on $\Delta \mu/\mu$,
this is inconsistent with claims of a smaller value of $\alpha$ from the many-multiplet 
method, unless fractional changes in $g_p$ are larger than those in $\alpha$ and $\mu$. 

\end{abstract}


\section{Introduction}
\label{sec:intro}

A critical assumption in the standard model of particle 
physics is that low-energy coupling constants and particle masses do not vary 
with space or time. This assumption breaks down in most theories that attempt to 
unify the standard model and general relativity (e.g. \citealp{marciano84}).  
The detection of such spatio-temporal variation in coupling constants like the fine 
structure constant $\alpha$, or the ratios of particle masses (e.g. the proton-electron
mass ratio $\mu \equiv m_p/m_e$), would imply new physics beyond the standard 
model and is hence of great interest (e.g. \citealp{uzan03}).

Comparisons between the redshifts of spectral lines detected in 
distant galaxies provide an important tool to probe changes in $\alpha$, 
$\mu$ and the proton gyromagnetic ratio $g_p$ over cosmological times 
(e.g. \citealp{wolfe76,dzuba99,chengalur03,flambaum07b}). Most such studies have 
yielded constraints on changes in $\alpha$, $\mu$ and $g_p$, with different 
systematic effects [see \citet{kanekar08b} for a recent review]. At present, the only 
technique that has found statistically-significant evidence for a change in 
one of the fundamental constants is the ``many-multiplet'' method \citep{dzuba99}. 
\citet{murphy04} obtained $\dal = (-5.7 \pm 1.1) \times 10^{-6}$ from Keck 
High Resolution Ultraviolet Echelle Spectrometer (HIRES) optical spectra of 
143~absorbers with a mean redshift $\langle z \rangle  = 1.75$, suggesting 
that $\alpha$ was smaller at earlier times. Other studies, applying a similar technique
to Very Large Telescope (VLT) data on smaller samples, have not confirmed this result 
(e.g. \citealp{levshakov06,srianand07b}). However, \citet{murphy08b} argue that 
the errors in these studies have been under-estimated, and that results from small 
samples are more prone to systematic effects related to the fitting of spectral 
components to complex absorption profiles. In the case of $\mu$, \citet{king08} used VLT 
Ultraviolet Echelle Spectrograph (UVES) spectra of ro-vibrational H$_2$ lines in 
three damped Ly$\alpha$ systems at $z \sim 2.6 - 3$ to find $\dmu = (-2.6 \pm 3.0) 
\times 10^{-6}$. Note that none of the above error estimates include recently-detected 
systematic effects due to distortions in the wavelength scales of the HIRES and UVES 
spectrographs (e.g. \citealt{griest10}). Finally, constraints on changes in 
$\alpha$, $\mu$ and $g_p$ have also been obtained from radio techniques (e.g. 
\citealp{carilli00,kanekar04b,kanekar05,murphy08}), although at lower redshifts.

Comparisons between the redshifts of \hi~21cm (hyperfine) and ultraviolet resonance 
dipole transitions are sensitive to changes in $X \equiv g_p \alpha^2/\mu$ \citep{wolfe76}. 
The best resonance lines for this method are those arising from {\it neutral} 
atomic species (e.g. \ci, Fe{\sc i}, etc), as these species are most likely to be 
physically associated with the \hi. The \ci\ multiplets are likely to 
be the best among the neutral resonance transitions as they typically arise in cold gas 
which also gives rise to the \hi~21cm absorption (e.g. \citealt{jenkins01,srianand05}). 
The ionization potentials of \ci\ and \hi\ are also similar, 11.3~eV for \ci\ 
and 13.6~eV for \hi\ (cf. Mg{\sc i}, which is more easily detectable than \ci\ 
in high-$z$ absorbers, has an ionization potential of 7.6~eV, as well as a high 
dielectronic recombination rate that can give significant Mg{\sc i} absorption 
in warm, ionized gas; \citealp{pettini77}). Finally, absorbers with a single 
(or dominant) spectral component in both \hi~21cm and \ci\ transitions are 
best-suited for such studies.

\hi~21cm and \ci\ absorption have hitherto both been detected in only one
high-$z$ absorber, the $z \sim 1.776$ system towards 1331+170, yielding 
a weak constraint on changes in $X \equiv g_p \alpha^2/\mu$ \citep{cowie95}. 
However, the \ci\ line in this absorber has two clear spectral components 
\citep{dessauges04}, implying ambiguities (and thus, large systematic errors) 
in the comparison with the \hi~21cm line. In this {\it Letter}, we report the 
detection of narrow, single-component \hi~21cm and \ci\ absorption in two absorbers 
at $z \sim 1.4-1.6$, that allow a high-sensitivity study of changes in 
the fundamental constants.

\begin{figure*}[t!]
\centering
\epsfig{file=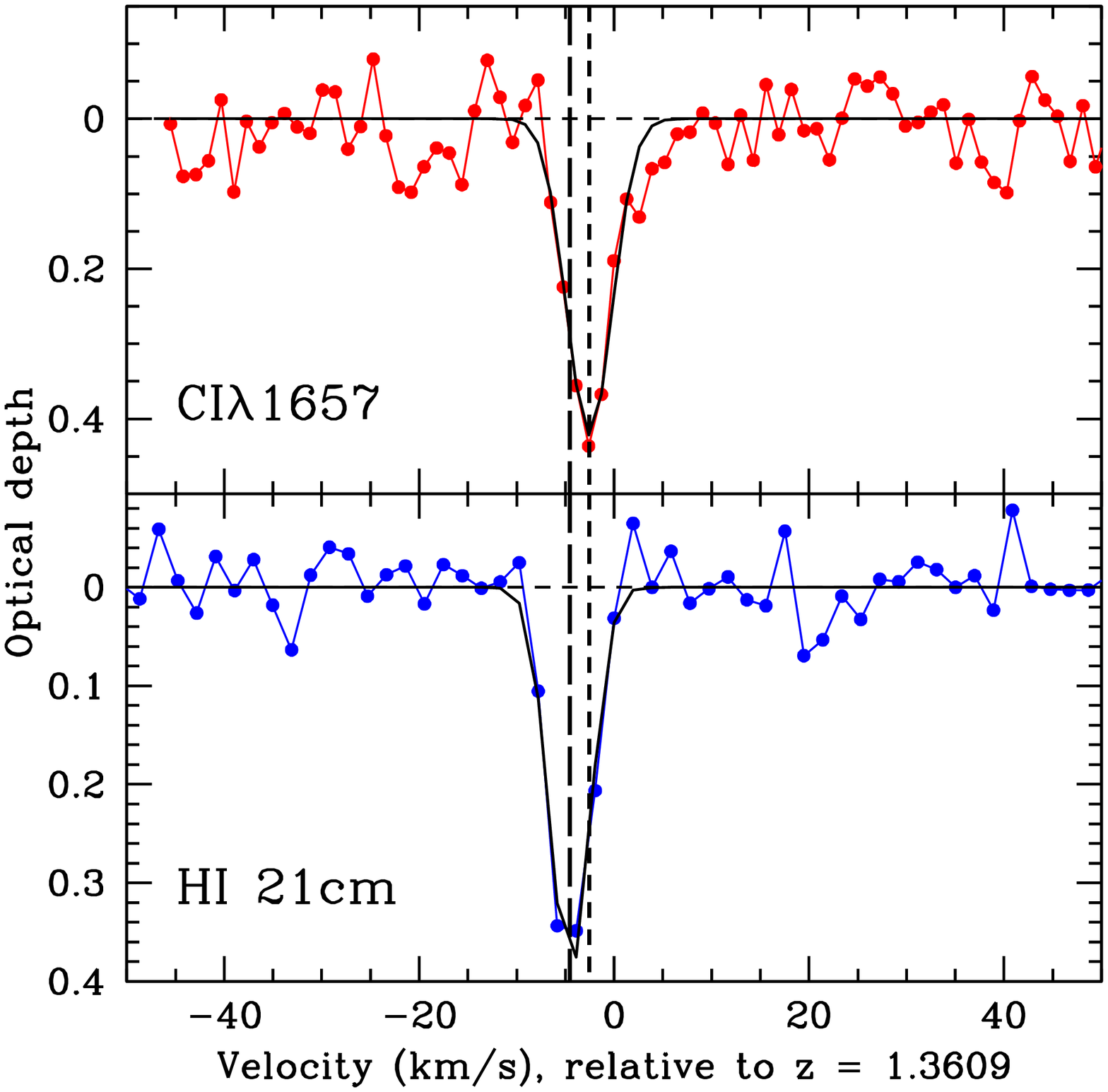,width=3.4in,height=3.4in}
\epsfig{file=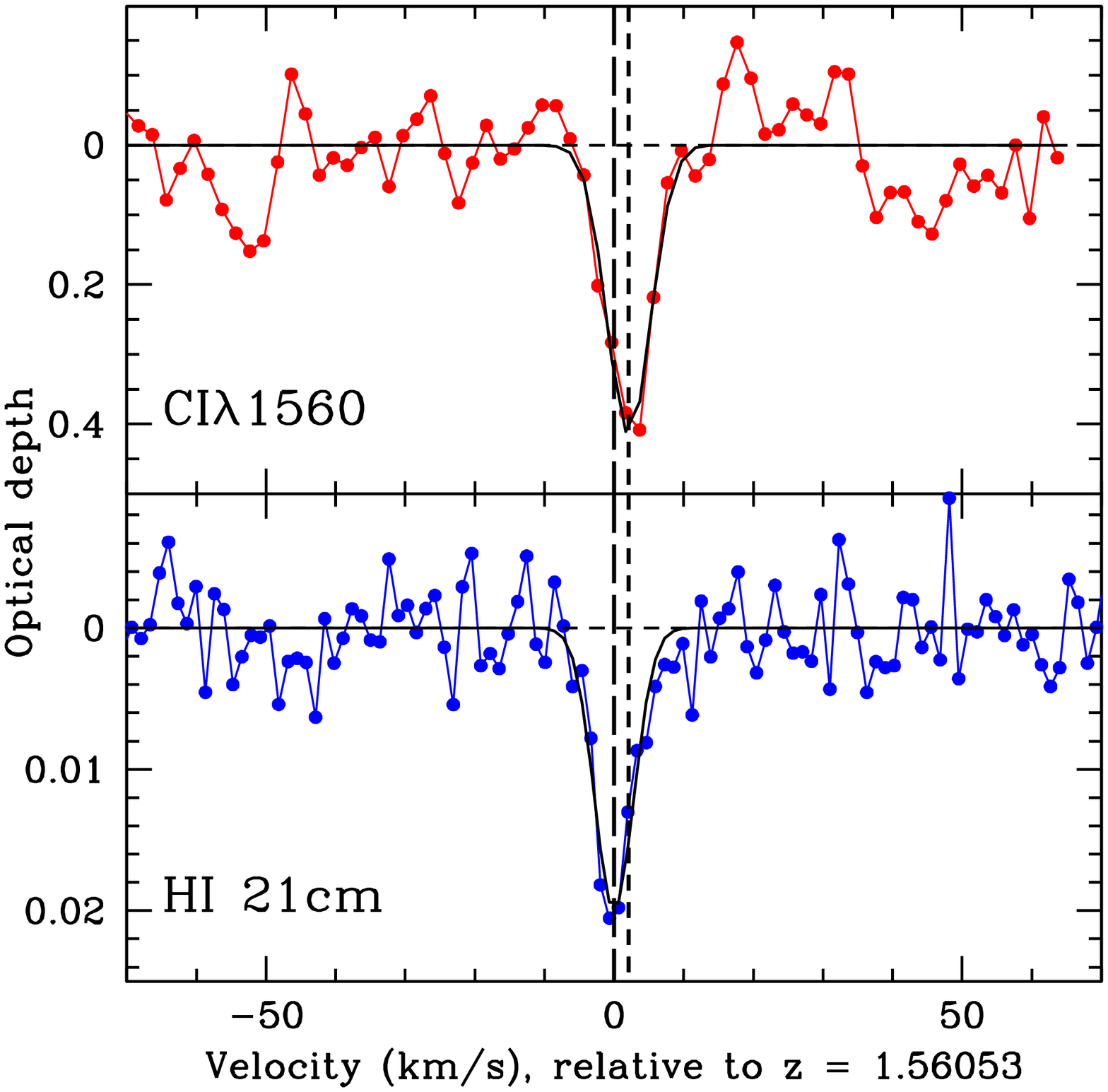,width=3.4in,height=3.4in}
\caption{\hi~21cm and \ci\ spectra towards [A]~Q2337$-$011 and [B]~Q0458$-$020. 
In each panel, the solid line shows the 1-gaussian fit to the spectrum. The 
small-dashed and large-dashed vertical lines indicate the \ci\ and \hi~21cm redshifts, 
respectively; the \ci\ redshift is seen to be higher than the \hi~21cm redshift in 
both panels.}
\label{fig:spectra}
\end{figure*}

\section{Spectra and results}
\label{sec:spectra}

The Giant Metrewave Radio Telescope (GMRT) 
and the Green Bank Telescope (GBT) were used to detect \hi~21cm absorption 
at $z \sim 1.3609$ towards Q2337$-$011 and $z \sim 1.5605$ towards Q0458$-$020, 
respectively, in a survey for \hi~21cm absorption in strong Mg{\sc ii} 
absorbers [GMRT proposals 7NKa02, 10NKa02, and GBT proposal 6A-026; see \citet{kanekar09b} for 
details]. The \hi~21cm spectra are shown in the lower panels of 
Figure~\ref{fig:spectra}. The root-mean-square (RMS) noise values, measured
from off-line regions in the spectra, are $\sim 0.03$ per 1.9~km/s 
channel (Q2337$-$011) and $\sim 0.0037$ per 1.3~km/s channel (Q0458$-$020), 
in optical depth units.

We then carried out a search for redshifted \ci\ absorption from the two absorbers, 
using the high-sensitivity Keck-HIRES spectrum of \citet{prochaska97} towards Q0458$-$020 
(observed on 31~October and 1~November 1995), and a new HIRES spectrum of Q2337$-$011 
(observed on 18 and 19~September 2006). This resulted in the detection of the \ci$\lambda$1560 
and \ci$\lambda$1657 transitions at $z \sim 1.3609$ towards Q2337$-$011, and the 
\ci$\lambda$1560 and \ci*$\lambda$1561 transitions at $z \sim 1.5605$ towards Q0458$-$020. 
Of these, the \ci$\lambda$1560 line towards Q2337$-$011 is blended with another 
line, while the \ci*$\lambda$1561 line towards Q0458$-$020 is detected at low 
significance; these will hence not be used in the later analysis. The \ci$\lambda$1657 
line towards Q2337$-$011 and the \ci$\lambda$1560 line towards Q0458$-$020 are shown 
in the upper panels of Figure~\ref{fig:spectra}. The signal-to-noise ratios per pixel are 
$\sim 25$ at a resolution of $R \sim 50000$ (Q2337$-$011), and $\sim 15$ at $R \sim 37000$ 
(Q0458$-$020), in the vicinity of the above \ci\ transitions.

For all spectra, the RMS noise was measured from absorption-free spectral 
regions around the line in question. A single-gaussian model was then used 
to independently fit each of the 1-D spectra in the \hi~21cm and \ci\ lines.
This model yielded an excellent fit to all spectra, with reduced chi-square values 
$< 1.06$ in all cases, and no evidence for statistically-significant features in 
the residual spectra. A Kolmogorov-Smirnov rank-1 test found all residual 
1-D spectra to be consistent (within $1.1\sigma$ significance) with being drawn 
from a normal distribution. The RMS noise values used for the spectral fits were 
scaled (marginally) to obtain $\chi^2 = 1$ in all cases; the resulting fits were then 
used to measure the peak absorption redshift for each transition. These 
redshifts are listed in columns~(2) and (3) of Table~\ref{table:fit}; note 
that the \ci$\lambda$1560 and \ci$\lambda$1657 lines have laboratory vacuum 
wavelengths of 1560.3092~\AA\ and 1657.9283~\AA, respectively \citep{morton03}, 
while the \hi~21cm line frequency is 1420.405751766~(1) MHz \citep{essen71}.

A useful test of the limiting accuracy in such fits can be obtained by 
combining the minimum uncertainties contributed by all line pixels to 
determine the best velocity accuracy that might be obtained with a given 
spectral fit \citep{murphy08b}. In all cases, the redshift errors in Table~\ref{table:fit} 
are larger than this limiting accuracy (by factors $\lesssim 2$), as expected 
for real spectra. We also tested that adding additional components to the fits 
does not significantly alter the results.

Assuming that the \hi~21cm and \ci\ lines arise in the same gas, the fractional 
change in $X \equiv g_p \alpha^2/\mu$ is related to the observed \hi~21cm and 
\ci\ redshifts by $\Delta X/X = [z_{\textrm{\ci}} -  z_{\textrm{21cm}}]/[1 + \bar{z}]$,
where $\bar{z} = (z_{\textrm{\ci}} -  z_{\textrm{21cm}})/2$ \citep{wolfe76}. 
This yields $\Delta X/X = (+6.64 \pm 0.84) \times 10^{-6}$ for the $z \sim 1.3609$ 
absorber towards Q2337$-$011 and $\Delta X/X = (+7.0 \pm 1.8) \times 10^{-6}$ 
for the $z \sim 1.5605$ absorber towards Q0458$-$020. Averaging these values 
(with equal weights) gives the result $\Delta X/X = (+6.8 \pm 1.0) \times 10^{-6}$, 
over $0 < z < 1.46$. We emphasize that the errors quoted here are purely statistical 
ones, from the fits; systematic effects are discussed in the next section.

\setcounter{table}{0}
\begin{table}
\begin{centering}
\caption{Parameters of the single-gaussian fits to the \hi~21cm and \ci\ spectra 
towards Q2337$-$011 and Q0458$-$020. }
\label{table:fit} 
\begin{tabular}{|c|c|c|c|c|}
\hline
	& $z_{\rm 21cm}$& $z_{\textrm{\ci}}$ & $\Delta V$$^\dagger$ & $\Delta X/X$ \\
 	&               &                    &  km/s      & $\times 10^{-6}$ \\            
\hline
&&& \\ 
Q2337$-$011 & 1.3608644 (13) & 1.3608801 (15) & $1.99 \pm 0.25$ & $+6.64 \pm 0.84$ \\
Q0458$-$020 & 1.5605300 (25) & 1.5605480 (38) & $2.11 \pm 0.53$ & $+7.0 \pm 1.8$   \\ 
&&&& \\
\hline
\end{tabular}
\end{centering}
$^\dagger$~The velocity offset of the \ci\ line from the \hi~21cm line.
\end{table}

\section{Systematic effects}
\label{sec:systematics}

\begin{figure}[t!]
\centering
\epsfig{file=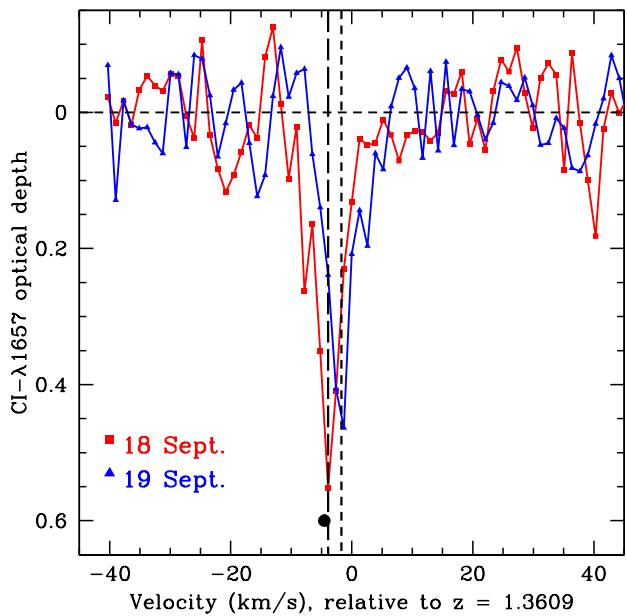,width=3.4in,height=3.4in}
\caption{A comparison between the \ci$\lambda$1657 line profiles obtained 
on 18 and 19~September, plotted as squares and triangles, respectively. The dashed vertical 
lines indicate the \ci\ redshifts obtained from 1-gaussian fits to each spectrum; these are 
offset from each other by $\sim 2$~km/s. The solid circle shows the \hi~21cm redshift; 
see main text for discussion.}
\label{fig:2337cI}
\end{figure}

At the outset, it should be emphasized that, unlike the many-multiplet method which only 
requires accurate {\it relative} wavelength calibration between different lines in the same 
optical spectrum, the comparison between \hi\ and \ci\ lines requires accurate 
{\it absolute} wavelength calibration of the optical spectra. On the other hand, the 
many-multiplet method is based on a first-order effect (the wavelengths of transitions used 
in this analysis have the same zeroth-order dependence on $\alpha$), while the present method 
uses the zeroth-order dependences of the hyperfine and resonance line frequencies 
on $\alpha$, $\mu$ and $g_p$. This implies that systematic effects (e.g. due to 
wavelength mis-calibration) are less important by an order of magnitude in the 
hyperfine/resonance comparison than in the many-multiplet analysis. For example,
a velocity uncertainty of $0.3$~km/s implies an error of $\dal = 10^{-5}$ in 
the many-multiplet method \citep{murphy01d}, but of $\Delta X/X = 10^{-6}$ 
in the hyperfine/resonance comparison.

The \ci\ lines in both absorbers are at {\it higher} redshifts than the \hi~21cm 
lines, by $\sim 2$~km/s. Systematic effects that might cause such a velocity offset 
include (1)~different relative isotopic abundances at $z \sim 1.5$ from the local 
Universe, that might cause the rest \ci\ wavelengths in the two absorbers to 
be higher than their laboratory values, (2)~errors in the absolute wavelength 
calibration of the optical spectra, and (3)~``local'' velocity offsets between 
the \ci\ and \hi~21cm lines in the two absorbers. Note that frequency 
mis-calibration of the \hi~21cm spectra is unlikely to be the source of 
such errors, as the absolute frequency scale here is set by the accuracy of 
masers and local oscillators (typically $< 10$~Hz, two orders of magnitude lower 
than the observed velocity offset between the \hi~21cm and \ci\ lines).

The velocity shifts of the $^{14}$C and $^{13}$C isotopic transitions relative 
to the main $^{12}$C isotopic lines are, respectively, $+0.46$~km/s and $+0.84$~km/s 
for the \ci$\lambda$1657 line, and $-3.29$~km/s and $-6.10$~km/s for the \ci$\lambda$1560 
line \citep{berengut06}. The isotopic velocity shift is clearly too small to account 
for the observed velocity offset in the case of the \ci$\lambda$1657 line. Conversely, 
in the case of the \ci$\lambda$1560 line, the isotopic shift yields the opposite sign 
from the observed velocity offset between \ci\ and \hi\ lines. We can thus rule out 
the hypothesis that differing carbon isotopic abundances (compared to Galactic values) 
might account for the observed velocity offset between the \hi~21cm and \ci\ transitions. 

Mis-calibration of the absolute wavelength scale of the optical spectra could
arise from a number of causes (e.g. \citealp{murphy01d}). In particular, 
the spectrum towards Q0458$-$020 was obtained in 1995, before Keck-HIRES
was fitted with an image rotator. It was thus not possible to hold the
slit perpendicular to the horizon during these observations, implying that 
atmospheric dispersion across the slit could produce errors in the wavelength 
scale \citep{murphy01d}. Further, while the data towards Q2337$-$011 were obtained 
in 2007, with the use of the image rotator, \citet{griest10} have found 
evidence for drifts in the Keck-HIRES wavelength 
scale with time, with amplitudes of $\sim 2$~km/s over multiple observing epochs.
The source of these velocity drifts is still unclear, although temperature 
changes, changes in the position of the quasar on the slit, as well as physical shifts 
in the echelle grating or cross-disperser are all possible causes.

To test for systematic effects in the absolute wavelength calibration of the 
optical spectra, we analysed archival Keck-HIRES data towards 0458$-$020 from
5 and 6~October~2004 (obtained using the image rotator). While this spectrum has
lower S/N than the 1995 spectrum (as the optically-variable quasar was in a fainter state 
in 2004), the unsaturated Ni{\sc ii}$\lambda$1317 transition from a higher-redshift ($z = 2.03945$) 
damped Lyman-$\alpha$ system is clearly detected in both spectra; this line lies at 
$\sim 4003$~\AA, within $10$~\AA\ of the \ci$\lambda$1560 line shown in 
Figure~\ref{fig:spectra}[B].  Cross-correlating the Ni{\sc ii} lines detected in 
the two HIRES spectra yielded a cross-correlation peak at a velocity offset of 
$+1.60 \pm 0.84$~km/s; the error was determined by cross-correlating 10000~pairs 
of simulated spectra with the shape and noise properties of the actual spectra. 
If this velocity offset between the Ni{\sc ii} lines arises due to atmospheric 
dispersion across the slit during the 1995 observations, it would imply that the 
\ci\ line redshift has been {\it under-estimated} in the 1995 spectrum. Correcting 
for this effect would {\it increase the redshift offset between the 
\hi~21cm and \ci\ lines}.

We also separately analysed the Keck-HIRES data towards Q2337$-$011 from 
18 and 19~September 2006, to test whether the same line redshift was obtained from 
the two runs. Figure~\ref{fig:2337cI} shows the \ci$\lambda$1657 line profiles 
obtained on 18 and 19~September; these are clearly offset from each other. 
A single-component gaussian fit was used to measure the \ci$\lambda$1657 redshift 
from each spectrum; the offset between the \ci$\lambda$1657 redshifts is $2.15 \pm 0.44$~km/s.
Similar velocity offsets ($\sim 2$~km/s) were seen between other lines in the two spectra. 
Note that a single ThAr lamp exposure was used to calibrate the two science exposures on Q2337$-$011 
on each day; the ThAr exposure was taken immediately before/after the science 
exposures. Interestingly, the \ci$\lambda$1657 redshift of 18~September [$z_\textrm{\ci} = 
1.3608694 (26)$] is in reasonable agreement with the \hi~21cm redshift 
[$z_\textrm{21cm} = 1.3608644 (13)$] , indicated by the solid circle in Figure~\ref{fig:2337cI}; 
the redshift offset between the \hi~21cm and \ci\ lines is dominated by the \ci\ data 
from 19~September. The offsets of $\sim 2$~km/s between our two observing epochs 
are consistent with the velocity drifts in the HIRES wavelength scale found by 
\citet{griest10} (see their Figure~5). We hence conclude that the absolute wavelength 
scale of HIRES could be in error by $\sim 2$~km/s. 

Finally, ``local'' velocity offsets between the \hi~21cm and \ci\ lines within the 
absorbing galaxies might also contribute to differences in the measured line redshifts. 
Both \ci\ and \hi~21cm absorption are expected to arise in cold gas, and, for both 
background quasars, a significant fraction of the radio flux density at the \hi~21cm line 
frequency arises from a compact core \citep{kanekar09b}. In fact, Q2337$-$011 has a 
highly inverted spectrum, indicating that most of the flux density arises from 
a self-absorbed core \citep{kanekar09b}. The cold cloud producing \ci\ absorption 
against the optical quasar is thus also likely to give rise to \hi~21cm absorption 
against the radio core. Assuming that gravity is important for cloud confinement 
on small scales, \citet{murphy03} obtain a velocity dispersion of $\sim 0.1$~km/s 
between different species in an individual cloud, which would yield a systematic error
of $\sim 3.3 \times 10^{-7}$ in $\Delta X/X$, significantly lower than our statistical
errors.  We note, however, that larger velocity offsets due to small-scale structure in 
the absorbing gas cannot formally be ruled out.

\section{Discussion}
\label{sec:discuss}

Prior to this work, the most sensitive result constraining changes in 
$X \equiv g_p \alpha^2/\mu$ using the hyperfine/resonance comparison was that of 
\citet{tzanavaris07}. These authors compared redshifts of the deepest absorption in 
\hi~21cm and low-ionization metal lines to obtain $\Delta X/X = (6.3 \pm 9.9) \times 
10^{-6}$ from a sample of nine absorbers at $0.23 < z < 2.35$. Note that these error 
estimate do not include systematic effects. Considering only statistical errors, 
our result, $\Delta X/X = [+6.8 \pm 1.0 (statistical)] \times 10^{-6}$ over 
$0 < \langle z \rangle  < 1.46$, is an order of magnitude more sensitive than that of 
\citet{tzanavaris07}. Further, the low-ionization metal lines used by 
\citet{tzanavaris07} could also arise in warm \hi\ or ionized gas, and are 
not necessarily associated with the cold \hi\ that gives rise to the 
\hi~21cm absorption. Most of the absorbers used by \citet{tzanavaris07} 
also have complex \hi~21cm and metal line profiles, and it is not necessary that 
the deepest absorption in the two types of transitions arises in the same spectral 
component \citep{kanekar06}. This could give systematic errors of $\gtrsim 10$~km/s, 
far larger than the statistical errors of \citet{tzanavaris07}, or the systematic 
errors in the present result.

The comparison between hyperfine and resonance transitions directly probes changes 
in $X \equiv g_p \alpha^2/\mu$, and one cannot obtain {\it independent} constraints 
on the individual constants without additional assumptions. Our result gives
$2\times\dal + [\Delta g_p/g_p] - \dmu = [+6.8 \pm 1.0 (statistical) \pm 6.7 (systematic)] 
\times 10^{-6}$. Note that $6.7 \times 10^{-6}$ is the {\it maximum} estimated error 
due to systematics in the absolute wavelength calibration of the optical spectra, 
(and not a $1 \sigma$ estimate, as in the case of the statistical error). This implies
that $2\times\dal + [\Delta g_p/g_p] - \dmu \ge [+0.1 \pm 1.0] \times 10^{-6}$.
The Keck/HIRES result of \citet{murphy04} is $\dal = (-5.7 \pm 1.1) \times 10^{-6}$, 
with $\langle z \rangle = 1.75$. Consistency between the two results would require 
either (1)~an additional wavelength calibration error of $\sim 1.7$~km/s in the \ci\ 
spectra, which appears unlikely, (2)~a ``local'' velocity offset of $\sim 1.7$~km/s 
(and with the appropriate sign) between the \hi\ and \ci\ lines in {\it both} absorbers, 
(3)~under-estimated errors in the many-multiplet result [e.g. the distortions in the 
wavelength scale found by \citet{griest10}], or (4)~$[\Delta g_p/g_p] - \dmu \ge (+1.14 \pm 0.24) 
\times 10^{-5}$ at $\langle z \rangle = 1.46$. In other words, assuming that the errors in 
one (or both) of the results have not been under-estimated, consistency between
the results requires that fractional changes in $\mu$ and/or $g_p$ are comparable
to those in $\alpha$. Strong constraints are available on fractional changes in $\mu$ at 
both higher and lower redshifts, $\dmu < 1.6 \times 10^{-6}$ at $z \sim 0.685$ \citep{murphy08} 
and $\dmu < 6.0 \times 10^{-6}$ at $z \sim 2.8$ \citep{king08}, making it unlikely that
$\dmu \sim 10^{-5}$ at $z \sim 1.46$. We thus conclude that the present result appears 
inconsistent with the smaller value of $\alpha$ at $\langle z \rangle \approx 1.75$ 
found by \citet{murphy04}, unless fractional changes in the proton gyromagnetic ratio $g_p$ 
are larger than those in $\alpha$ and $\mu$. 

In summary, we have detected narrow \hi~21cm and \ci\ absorption in two absorbers at 
$z \sim 1.4-1.6$ towards Q0458$-$020 and Q2337$-$011, using the Keck telescope, 
the GMRT and the GBT. The \ci\ and \hi~21cm line frequencies have different dependences 
on the fundamental constants $\alpha$, $\mu \equiv m_p/m_e$ and $g_p$, allowing us 
to use the measured \ci\ and \hi~21cm line redshifts to test for putative changes 
in these constants. Comparing the \hi~21cm and \ci\ redshifts in the two absorbers yields 
the result $\Delta X/X = [+6.8 \pm 1.0 (statistical) \pm 6.7 (systematic)] \times 10^{-6}$ over 
$0 < \langle z \rangle 1.46$, where $X \equiv g_p \alpha^2/\mu$. This is inconsistent
with evidence for a smaller value of $\alpha$ at similar redshifts from the many-multiplet 
method, unless fractional changes in $g_p$ are larger than those in $\alpha$ and $\mu$.
Systematic errors in the hyperfine/resonance comparison are currently dominated by errors 
($\sim 2$~km/s) in the absolute wavelength calibration of the optical spectra.
However, the comparison between \hi~21cm and \ci\ lines has a high sensitivity,
and it should be possible to significantly reduce systematic effects by the use
of new calibration techniques (e.g. laser frequency combs; \citealt{steinmetz08}).
Increasing the number of detections of redshifted, narrow \hi~21cm and \ci\ absorption 
is thus of much importance.

\acknowledgments
Some of the data presented herein were obtained at the W.M. Keck Observatory, 
which is operated as a scientific partnership among the California Institute 
of Technology, the University of California and the National Aeronautics and 
Space Administration. The Observatory was made possible by the generous 
financial support of the W.M. Keck Foundation. We thank the staff of the GMRT 
and the GBT for help during these observations. The GMRT is run by the National 
Centre for Radio Astrophysics of the Tata Institute of Fundamental Research. 
The National Radio Astronomy Observatory is operated by Associated Universities, 
Inc, under cooperative agreement with the NSF. NK acknowledges support from the 
Max-Planck Society, and from the Department of Science and Technology via 
a Ramanujan Fellowship. J.X.P. acknowledges 
funding through NSF CAREER grant (AST--0548180) and NSF grant (AST--0709235).

\bibliographystyle{apj}

\end{document}